\RequirePackage{etoolbox}
\let\originalshowhyphens\showhyphens

\documentclass[sigconf]{acmart}

\let\showhyphens\originalshowhyphens

\usepackage{graphicx}
\usepackage{booktabs}
\usepackage{tcolorbox}
\usepackage{multirow}
\usepackage{siunitx}
\usepackage{url}
\usepackage[table]{xcolor}

\tcbset{
    mysummarybox/.style={
        colback=blue!5!white, 
        boxrule=0pt,          
        colframe=blue!5!white,
        sharp corners,        
        boxsep=5pt,           
        left=0pt, right=0pt, top=0pt, bottom=0pt, 
        fontupper=\normalsize 
    }
}

\usepackage{enumitem}

\newlist{rquestions}{enumerate}{1}
\setlist[rquestions]{
    label=\textbf{RQ\arabic*:},  
    font=\itshape,               
    align=left,                  
    labelwidth=!,                
    labelsep=0.5em,              
    leftmargin=*,                
}

\copyrightyear{2026}
\acmYear{2026}
\acmConference[MSR '26]{23rd International Conference on Mining Software Repositories}{April 13--14, 2026}{Rio de Janeiro, Brazil}
\acmBooktitle{23rd International Conference on Mining Software Repositories, April 13--14, 2026, Rio de Janeiro, Brazil}
\acmPrice{}
\acmDOI{}

\begin{document}

\title{Code Change Characteristics and Description Alignment: A Comparative Study of Agentic versus Human Pull Requests}

\author{Dung Pham}
\affiliation{
  \department{Department of Computer Science} 
  \institution{Trent University}
  \city{Peterborough}
  \state{Ontario}
  \country{Canada}}
\email{dungpham290198@gmail.com}

\author{Taher A. Ghaleb}
\orcid{0000-0001-9336-7298}
\affiliation{
  \department{Department of Computer Science} 
  \institution{Trent University}
  \city{Peterborough}
  \state{Ontario}
  \country{Canada}}
\email{taherghaleb@trentu.ca}

\begin{abstract}
AI coding agents can autonomously generate pull requests (PRs), yet little is known about how their contributions compare to those of humans. We analyze 33,596 agent-generated PRs (APRs) and 6,618 human PRs (HPRs) to compare code-change characteristics and message quality.  
We observe that APR-introduced symbols (functions and classes) are removed much sooner than those in HPRs (median time to removal 3 vs. 34 days) and are also removed more often (symbol churn 7.33\% vs. 4.10\%), reflecting a focus on other tasks like documentation and test updates.
Agents generate stronger commit-level messages (semantic similarity 0.72 vs. 0.68) but lag humans at PR-level summarization (PR–commit similarity 0.86 vs. 0.88). Commit message length is the best predictor of description quality, indicating reliance on individual commits over full-PR reasoning.
These findings highlight a gap between agents' micro-level precision and macro-level communication, suggesting opportunities to improve agent-driven development workflows.
\end{abstract}




\maketitle

\section{Introduction}
Large language models are increasingly used for software development
tasks such as code generation and pull request (PR) creation
\cite{moautopr, watanabe2025use, cotroneo2025human}. Recent agentic systems extend these capabilities through multi-step contributions \cite{wang2025ai},
motivating an examination of how agent-generated pull requests differ from human ones in code change patterns and communication quality, which
directly affect reliability, maintainability, and security \cite{cotroneo2025human}. PRs communicate changes through descriptions
and commit messages. Automated commit message generation still exhibits
poor generalization ability \cite{wu2025empirical}. Since PRs often span
multiple commits, generating PR-level descriptions requires additional
effort \cite{sakib2024automatic}. This challenge underscores the need to
understand whether agent-generated PRs (APRs) differ from human PRs (HPRs) in their
communication effectiveness.

In this paper, we address two research questions: 
\textbf{(RQ1)} \textit{How do APRs and HPRs differ in code change characteristics (files changed, code churn, lines added/removed, and change purposes)?}
\textbf{(RQ2)} \textit{How well do APR descriptions and commit messages align with code changes?}
To do this, we introduce three alignment metrics: PR--commit similarity, patch--message similarity, and an LLM-based consistency score, and build a classification model to identify the factors that predict strong agent-generated PR descriptions.
Our results reveal that APRs consistently make smaller, more focused changes and excel at describing them at the commit level, but they perform worse than humans when articulating the broader, PR-level narrative.

\vspace{3pt}
\noindent\textbf{Paper organization.} Section \ref{sec:empirical} describes our data collection. Sections \ref{sec:code_change_characteristics} and \ref{sec:narrative_coherence} analyze code change characteristics (RQ1) and description alignment (RQ2). Section \ref{sec:implications} discusses the implications, Section \ref{sec:threats_to_validity} discusses threats to validity, Section \ref{sec:related_works} reviews related work, and Section \ref{sec:conclusion} concludes and outlines future work.

\vspace{3pt}
\noindent\textbf{Replication Package: } Our data,
scripts, notebooks, prompts and results can be found in our replication package \cite{replication_package}

\section{Empirical Analysis and Results}
\label{sec:empirical}
\subsection{Data collection}
In this study, we use the AIDev-pop subset of the AIDev dataset (Oct 28, 2025 update)~\cite{li2025aidev}, which contains 33,596 curated agentic PRs (APRs) from five major autonomous coding agents:
OpenAI Codex (21,799 PRs), Devin (4,827 PRs), GitHub Copilot (4,970 PRs), Cursor (1,541 PRs), and Claude Code (459 PRs). We also use a dataset of 6,618 human PRs (HPRs), and because detailed commit information was not provided, we constructed a complementary dataset by retrieving commit IDs, changed files, change types, patch diffs (added and removed lines of code), and commit messages via the GitHub API. The APRs span 2024-12 to 2025-07 and the HPRs 2025-01 to 2025-06, covering comparable periods.
We further computed benchmark metrics (RQ2) from samples of the two datasets, by randomly sampling 663 instances (corresponding to a 99\% confidence level with a $\pm$5\% margin of error) from the cleaned training set (138,895 records) of Tire et al.~\cite{tire2025evaluating} and 664 instances from the 249,688 test samples of the CommitBench dataset~\cite{schall2024commitbench}.

\subsection{Terminologies}
We define agentic PRs as \textbf{APRs} and human PRs as \textbf{HPRs}. The \textbf{Merged PR ratio} is computed over all PRs with a close date: PRs with a merge date are considered successful, others are not. \textbf{Task type} refers PR purpose, e.g., feature (\textit{feat}) or bug fix (\textit{fix}) as in Hao Li et al.~\cite{li2025aidev}. A \textbf{Directory} is the immediate parent folder of a changed file. \textbf{Symbols} are functions or classes extracted from patch diffs. \textbf{Symbol churn} refers to symbols that are introduced in a commit and subsequently removed in a later commit. We measure \textbf{PR--Commit Similarity} (semantic alignment between PR description and commit messages) and \textbf{Patch--Commit Similarity} (alignment between diff and messages). The \textbf{LLM-based Consistency Score} is a GPT-4o rating of the quality of the commit message.

\subsection{Code Change Characteristics}
\label{sec:code_change_characteristics}
\vspace{-1pt}
It is important to understand how agentic code changes differ from human changes, to help teams identify tasks agents handle well, where they fall short, and what needs extra review before merging. 

\vspace{4pt}
\noindent\textbf{\large RQ1.1: Do agents and humans differ in the footprint of their code modifications?}

\vspace{2pt}
\noindent\textbf{Approach.}
We compute the merge rate and four median per-PR change metrics (commits, changed files, directories, and lines) for 31,268 closed APRs and compare them with 6,135 closed HPRs. We model merge outcomes using multivariate logistic regression with footprint metrics (commits, files, directories) to assess their relation to merge success. We model PR task types with Poisson regression on the footprint, chosen for interpretability and to quantify effect sizes via odds ratios (OR) or incidence rate ratios (IRR), with statistical significance tested using $p$-values. Poisson regression is appropriate for non-negative, often non-normal and positively skewed count data~\cite{hutchinson2005analysis}. We assess statistical differences between APRs and HPRs in these metrics are evaluated using the Mann--Whitney U test~\cite{mann1947test} and proportional z-test~\cite{agresti2011categorical}, with $\alpha = 0.05$.

\vspace{2pt}
\noindent\textbf{Results.}
\textbf{\emph{APRs show lower merge rates and smaller change footprints, and a higher number of commits in a PR is associated with reduced merge success.}}
Table~\ref{tab:pr_metrics} shows that APRs have a lower merge rate than HPRs (76.80\% vs.\ 82.82\%, $p < 0.001$) and smaller footprints: fewer commits (median 1 vs.\ 2, $p < 0.001$), files (median 3 vs.\ 4, $p < 0.001$), and directories (median 1 vs.\ 2, $p < 0.001$). The number of changed lines does not differ significantly ($p = 0.131$). Logistic regression indicates that the number of commits is a strong negative predictor of merging (OR = 0.95, $p < 0.001$), with each additional commit reducing merge odds by about 5\%; the numbers of files and directories changed have no significant association ($p > 0.01$).

This lower number-of-commits pattern is associated with PR task types. The Poisson regression shows that task type is significantly associated with commit count. We use feature development (\textit{feat}) as the reference group, since it is most common and represents a typical PR. Documentation (IRR = 0.74, $p < 0.001$), test (IRR = 0.72, $p < 0.001$), and style (IRR = 0.64, $p < 0.001$) PRs involve about 26\%, 28\%, and 36\% fewer commits than feature PRs. In contrast, chore (IRR = 1.13, $p < 0.001$) and refactoring (IRR = 1.15, $p < 0.001$) PRs involve about 13\% and 15\% more commits, respectively, compared to the reference group. Table~\ref{tab:task_types_topk} shows that feature (feat) and bug fix (fix) tasks are the two most common task types for both groups, but the third and fourth most frequent tasks differ: APRs focus more on documentation and test tasks, whereas HPRs handle more build and chore tasks. This helps explain why APRs have fewer commits per PR than HPRs. Future research should further investigate possible causes, such as task selection bias, agent prompting strategies, or autonomy levels, to derive more actionable guidance.

\begin{table}[t]
    \centering
    \caption{Agentic PRs (APR) vs. Human PRs (HPR) Metrics.}
    \vspace{-8pt}
    \label{tab:pr_metrics}
    \begin{tabular}{p{5.6cm}rr}
        \toprule
        \textbf{Metric} & \textbf{APR} & \textbf{HPR} \\
        \midrule
        \# of PRs                        & 31,268  & 6,135 \\
        Merge Rate                       & 76.80\% & 82.82\% \\
        Median \# of Commits             & 1       & 2 \\
        Median \# of Files Changed       & 3       & 4 \\
        Median \# of Directories Changed & 1       & 2 \\
        Median \# of Lines Changed       & 90      & 88 \\
        \bottomrule
    \end{tabular}
    \vspace{-9pt}
\end{table}

\begin{table}[ht]
    \centering
    \vspace{-4pt}
    \caption{Top Task Types for APR and HPR}
    \vspace{-8pt}
    \label{tab:task_types_topk}
    \begin{tabular}{p{1.3cm}p{3cm}p{3cm}}
        \toprule
        \textbf{Rank} & \textbf{APR: Top Tasks} & \textbf{HPR: Top Tasks} \\
        \midrule
        Top 1 & Feat (42.79\%)  & Feat  (28.48\%) \\
        Top 2 & Fix  (23.59\%)  & fix   (27.19\%) \\
        Top 3 & Docs (11.95\%)  & Chore (13.32\%) \\
        Top 4 & Test (7.22\%)   & Build (9.60\%)  \\
        \bottomrule
    \end{tabular}
    \vspace{-7pt}
\end{table}

\vspace{3pt}
\noindent\textbf{\large RQ1.2: Do agents and humans differ in symbol churn and the timing of subsequent modifications?}

\vspace{2pt}
\noindent\textbf{Approach.}
We evaluate PR quality using two metrics: (1) time to next modification and (2) symbol churn. For time to next modification, we track when each file is first added and merged, then compute the number of days until its next modification in a later PR. For symbol churn, we extract functions and classes from patch diffs for the top three languages in the AIDev-Pop dataset (TypeScript, Python, and Go)~\cite{li2025aidev} using regular-expression patterns, then measure the days from symbol introduction to first removal. Differences between APRs and HPRs are assessed with the Mann--Whitney U test for days to modification and churn, and a proportional z-test for modification and symbol-churn ratios (all at $\alpha = 0.05$).

\vspace{2pt}
\noindent\textbf{Results.}
\textbf{\emph{Files and symbols introduced by APRs change more frequently and much sooner than those in HPRs.}} 
Table~\ref{tab:code_quality} summarizes the key findings on time to next modification and symbol churn. APRs show a higher file-modification ratio than HPRs (14.36\% vs. 3.36\%, $p < 0.001$) and are modified far earlier (median 0.88 days vs. 23 days, $p < 0.001$). For symbol-level changes, the pattern is similar: among 7{,}338 APR symbols and 8{,}089 HPR symbols, APRs exhibit a higher churn rate (7.33\% vs. 4.10\%, $p < 0.001$) and their symbols are removed much sooner (median 3 days vs. 34 days, $p < 0.001$).
This early churn can be explained with the task-type distribution observed in RQ1.1: For APRs, documentation and tests are among the top categories of work (table~\ref{tab:task_types_topk}), areas that naturally evolve quickly. For example, in one documentation-related PR\footnote{\href{https://github.com/Azure/azure-sdk-for-python/pull/41463/commits/a76fdf5f713f043d825f401272e8e69598e07e4b\#diff-b1b0d908b2f008886da37adad4d4780a4b2f4f94f677b09fb76afa2218748f43}{https://github.com/Azure/azure-sdk-for-python/pull/41463}}, the function \texttt{send\_with\_run\_in\_executor} was introduced and then removed just two days later in a subsequent PR\footnote{\href{https://github.com/Azure/azure-sdk-for-python/pull/41352/commits/10714e1853532e9ae738d0913a2aaa870e65bc58\#diff-b1b0d908b2f008886da37adad4d4780a4b2f4f94f677b09fb76afa2218748f43}{https://github.com/Azure/azure-sdk-for-python/pull/41352}}.

\begin{table*}[t]
    \centering
    \caption{Time to Next Modification and Code Churn Comparison between APR and HPR}
    \vspace{-10pt}
    \label{tab:code_quality}
    \resizebox{0.92\textwidth}{!}{
    \begin{tabular}{l r r r r r r}
        \toprule
        & \multicolumn{3}{c}{\textbf{Time to Next Modification}} & \multicolumn{3}{c}{\textbf{Code Churn}} \\
        \cmidrule(lr){2-4} \cmidrule(lr){5-7}
        \textbf{Type} & \textbf{\#New Added Files} & \textbf{\% Modification} & \textbf{Median \#Days} & \textbf{\#New Functions/Classes} & \textbf{\%Churn Rate} & \textbf{Median \#Days} \\
        \midrule
        APR & 87,277 & 14.36\% & 0.88 & 7,338 & 7.33\% & 3 \\
        HPR & 23,989 & 3.36\% & 29.2 & 8,089 & 4.10\% & 34 \\
        \bottomrule
    \end{tabular}
    }
    \vspace{-6pt}
\end{table*}

\subsection{Description Alignment}
\label{sec:narrative_coherence}
\vspace{-2pt}
It is important to examine how well agents’ PR descriptions reflect the underlying code changes. Understanding description quality and its drivers helps reveal where agents fall short and how to improve PR clarity and reliability.

\vspace{4pt}
\noindent\textbf{\large RQ2.1: How well do PR descriptions and commit messages align with code changes?}

\vspace{3pt}
\noindent\textbf{Approach.}
We address this RQ using three alignment metrics:  
\textbf{(1) PR–Commit Similarity:} We measure the semantic similarity between PR descriptions and commit messages by using GistEmbed~\cite{solatorio2024gistembed}, the top-performing model in prior benchmarks~\cite{sutriawan2025ritzkal}, by computing embeddings for both texts and then calculating cosine similarity.
\textbf{(2) Patch–Commit Similarity:} We use a fine-tuned CodeT5-Without-History model~\cite{eliseeva2023commit} to generate commit messages from patch diffs, as it is recognized as a state-of-the-art model for commit message generation~\cite{zhang2024rag} and is widely used as a baseline in prior work~\cite{shi2022race,wu2025empirical}. Using the history-free variant isolates patch-level semantic alignment and avoids stylistic or project-specific confounds. We then embed both generated and actual messages with GistEmbed and compute cosine similarity.
\textbf{(3) LLM Consistency Score:} We use GPT-4o to evaluate the metric based on established criteria for good commit messages~\cite{tian2022makes}, as it aligns well with human judgments, outperforms traditional lexical metrics~\cite{liu2023g}, and has been widely adopted as a judge model in LLM-as-a-Judge frameworks~\cite{huang2025empirical, wei2024systematic,mohammadkhani2025checklist}. Full prompt is in our replication package~\cite{replication_package}. This RQ focuses on assessing description alignment, so deleted symbols are retained to preserve generalizability.

We filter PR text for English, as it constitutes the vast majority of the dataset ($\approx$98\%), and to ensure linguistic consistency, using the XLM-R model~\cite{papluca_xlm-roberta-base-language-detection_2022} with a confidence threshold $>0.7$. XLM-R has strong performance on language identification tasks~\cite{conneau2020unsupervised}, yielding 22{,}633 APRs and 4{,}009 HPRs. From these, we sample 647 APRs and 571 HPRs (99\% confidence level, 5\% margin of error).
Due to the lack of ground-truth labels for PR--code alignment, we use high-quality datasets from prior work as benchmarks: 663 PR--commit instances from Tire et al.~\cite{tire2025evaluating} and 664 patch--commit instances from CommitBench~\cite{schall2024commitbench}. Statistical significance is assessed using the Kruskal--Wallis test~\cite{kruskal1952use} followed by Dunn’s post-hoc test~\cite{dunn1964multiple}. We report effect sizes using epsilon-squared ($\varepsilon^2$).

\begin{figure}[ht]
    \centering
    \includegraphics[width=\linewidth]{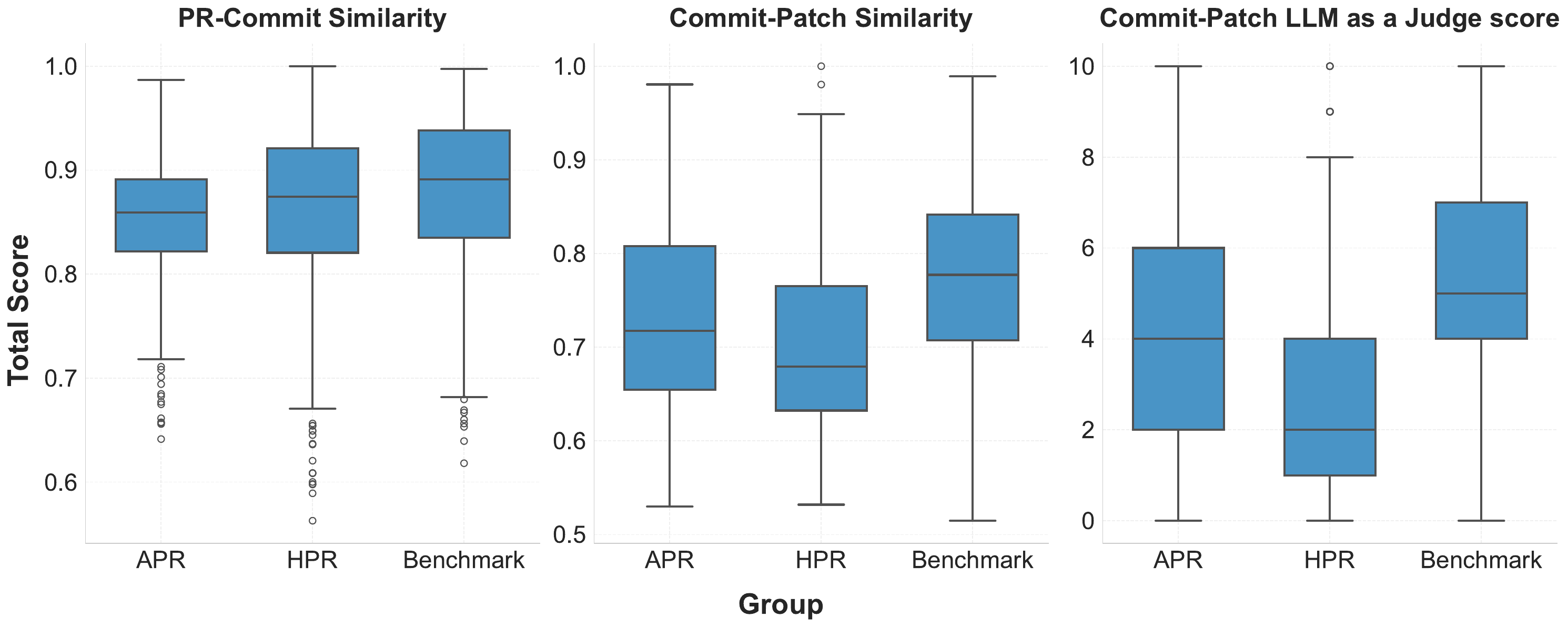}
    \vspace{-20pt}
    \caption{Distribution of PR-commit similarity score, Patch-commit similarity score and LLM-based consistency score.}
    \Description{Three boxplots comparing Agentic PRs (APR) and Human PRs (HPR). 1) PR-Commit Similarity: HPR median is slightly higher than APR. 2) Patch-Commit Similarity: APR median is higher than HPR. 3) LLM Score: APR median is higher (4) compared to HPR (2).}
    \label{fig:box_plot_2.1}
    \vspace{-10pt}
\end{figure}

\vspace{3pt}
\noindent\textbf{Results.}
\textbf{\emph{Agents are better at describing individual commits, while humans are better at summarizing PR-level changes.}} 
As shown in Figure~\ref{fig:box_plot_2.1}, APRs achieve higher commit-level alignment than HPRs both in semantic similarity (median 0.72 vs.\ 0.68, $p < 0.001$) and in LLM-as-a-judge scores (median 4/10 vs.\ 2/10), with $p < 0.001$ and effect sizes of $\varepsilon^2 \approx 0.12$ and $0.17$, respectively. In contrast, at the PR level, APRs score lower than humans in PR–commit similarity (0.86 vs.\ 0.88, $p < 0.001$), but with an effect size of $\varepsilon^2 \approx 0.04$. All differences are statistically significant under the Kruskal–Wallis test with Dunn’s post-hoc correction ($\alpha = 0.05$).

These results show that agents produce precise commit-level messages, but struggle to combine multiple commits into a coherent PR-level summary, which humans perform better. Benchmark results also show that both APRs and HPRs still lag behind high-quality references, leaving room for improvement at both levels.

\vspace{4pt}
\noindent\textbf{\large RQ2.2: What factors predict APR description quality?}

\vspace{2pt}
\noindent\textbf{Approach.}
To study what distinguishes strong agent-generated PR descriptions from weak ones, we frame the problem as a binary classification task. A description is labeled ``good'' if its PR–commit similarity score is $\ge 0.9$. Following prior work~\cite{ortiz2025transformer}, which notes that semantic-similarity thresholds should be tuned to application goals, we select 0.9 based on the distribution of PR--commit similarity scores of (Figure~\ref{fig:box_plot_2.1}), where it roughly corresponds to the 75th percentile and thus marks the cutoff for the top 25\% highest-scoring descriptions.
We then model this classification task using Random Forest (RF), k-Nearest Neighbors (KNN), Light Gradient Boosting Machine (LightGBM), and Extreme Gradient Boosting (XGBoost) classifiers. We also perform a sensitivity analysis to examine the impact of varying the threshold from the median to the 75th percentile (0.85--0.9 in increments of 0.01) on feature importance.
We construct three categories of features:
\textbf{(1) Descriptive attributes:} title length, body length, task type, template score, code-snippet flag;
\textbf{(2) Development activity:} related-issue count, commit count, average commit-message length; and
\textbf{(3) Code changes:} number of files changed, additions, deletions, total changes, change-status distribution, directories changed.
We split the data into 80\% training ($17,728$ PRs) and 20\% testing ($4,433$ PRs). Last, we apply SHAP to identify the strongest features to predict high-quality descriptions.

\vspace{2pt}
\noindent\textbf{Results.}
\textbf{\emph{Commit message length is the strongest predictor of a good APR description (max SHAP: 0.57)}}, followed by PR title and PR body length (max SHAP: 0.21 and 0.20, respectively). As shown in Figure~2, high-quality APR descriptions are strongly associated with commit messages $\ge$ 13 words, PR titles $\ge$ 7 words, and PR bodies $\ge$ 80 words.
We observe that RF outperformed other classifiers across all metrics (Accuracy: 86.22\%, AUC: 79.86\%, Precision: 72.8\%, Recall: 31.9\%, F1-score: 44.40\%) and was therefore selected for SHAP analysis. The relatively low F1-score results from the strict PR--commit similarity threshold ($\ge 0.9$) defining high-quality descriptions, creating strong class imbalance and favoring precision over recall. Threshold sensitivity analysis (0.85--0.9) shows F1-score decreasing from $\approx$71\% to 44\% as the cutoff increases, while AUC rises from $\approx$73\% to 79\%. As our goal is feature attribution rather than deployment-ready classification, RF is sufficient to identify the dominant predictors of high-quality descriptions.

\begin{figure}[t]
    \centering
    \vspace{-5pt}
    \includegraphics[width=\linewidth]{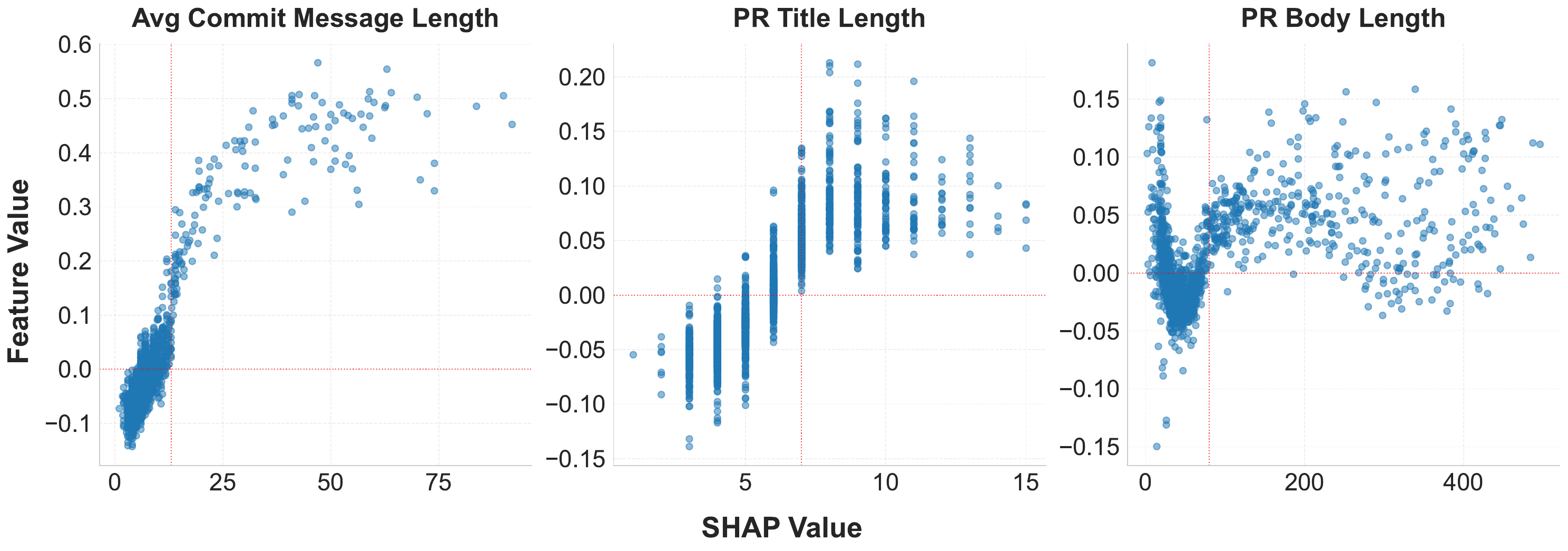}
    \vspace{-18pt}
    \caption{SHAP values of the top 3 features that contributes to defining a good PR description}
    \Description{SHAP dependence plots for the top three features. 1) Average Commit Message Length shows positive SHAP values (positive impact) when the length exceeds approximately 13 words. 2) PR Title Length shows a positive impact starting at around 7 words. 3) PR Body Length shows a positive contribution when the word count is above 80 words.}
    \label{fig:shap_value_by_feature}
    \vspace{-11pt}
\end{figure}

\section{Implications}
\label{sec:implications}
\vspace{-1pt}
Our findings offer the following implications:

\vspace{2pt}
\noindent\textbf{Project maintainers.} Assign AI agents narrowly scoped tasks to avoid unnecessary rework and prevent the merge-rate drop seen with larger changes. Carefully review newly introduced symbols, which are prone to high churn, and use length-based heuristics to flag brief or low-quality PR descriptions for extra review.

\vspace{2pt}
\noindent\textbf{AI Agent builders.} Move beyond simply aggregating commit messages and train agents to reason over the full changeset, enabling the generation of more coherent and holistic PR descriptions that better capture the intent, scope, and rationale of changes.

\vspace{2pt}
\noindent\textbf{Researchers.} Investigate methods to better evaluate PR quality and agent behavior, including developing metrics that capture semantic alignment, reviewing agent-specific differences, and exploring cross-language generalization of code change patterns.

\section{Threats to validity}
\label{sec:threats_to_validity}
\vspace{-1pt}

\noindent\textbf{Construct Validity.} We use semantic similarity and LLM-based scores as proxies for PR-description quality, which may not fully capture factual correctness and may reflect LLM self-preference bias. We mitigate this by combining multiple alignment metrics and manually verifying a small sample. We also use CodeT5-generated commit messages as a proxy for patch semantics; while not ideal, this remains a reasonable approximation given the benchmark group’s high similarity scores and the use of an LLM-as-a-judge metric. The ``good'' description threshold ($\ge 0.9$) is heuristic, but feature importance remained consistent across sensitivity tests. The model’s 31.9\% recall shows it misses most good descriptions, indicating that, despite its feature-attribution intent, the identified predictors do not fully capture what makes a description good.

\vspace{2pt}
\noindent\textbf{Internal Validity.} Our symbol-churn analysis (RQ1.2) uses regular expressions (regex) to extract function and class definitions for TypeScript, Python, and Go. We rely on regex rather than AST-based methods like Tree-sitter~\cite{tree_sitter} because patches contain limited code context~\cite{xiong2025contextual}, though the code is expected to be syntactically valid~\cite{gong2024ast}. To enhance reliability, we focus on the three languages with clearly recognizable tokens (\texttt{def}, \texttt{function}, \texttt{func}, \texttt{class}). Regex may miss symbols or capture false positives, e.g., commented or nested code, so we restrict the analysis to languages where extraction is most reliable. Manual inspection of a small sample further supports this choice, showing minimal false positives.

\vspace{2pt}
\noindent\textbf{External Validity.} Agent capabilities evolve rapidly, so our results represent a snapshot of current model performance and may change as agents improve reasoning and long-context understanding. Additionally, our churn analysis focuses on dynamic languages and may not generalize to statically typed ecosystems, e.g., Java and C++, where type constraints, annotations, generics, and other language-specific constructs complicate symbol extraction from partial diffs, preventing reliable regex-based parsing.

\section{Related works}
\label{sec:related_works}
\vspace{-1.5pt}
The closest prior work is Watanabe et al.~\cite{watanabe2025use}, who compared HPRs and APRs of one agent (Claude Code). We study five agents and add measures such as symbol churn and time to next modification. Unlike Watanabe et al., who report APRs introduce more code, we find no significant difference in lines changed. Both studies find feature development (\textit{feat}) and bug fixes (\textit{fix}) dominate, with documentation and tests more frequent in APRs.
Zi et al.~\cite{zi2025agentpack} introduced the AgentPack dataset, showing that co-authored changes are narrowly scoped and that LLM-generated commit messages capture intent well. However, they do not assess merge success or symbol stability and use only message length as a quality metric; our work fills these gaps with multiple quality metrics.

Prior PR-description research has focused on generating descriptions~\cite{sakib2024automatic,tire2025evaluating} or commit messages~\cite{eliseeva2023commit}, not alignment with code changes. We introduce metrics quantifying consistency between descriptions and code modifications for humans and agents, and model predictors of high-quality descriptions. Related work on commit message faithfulness~\cite{liu2025hallucinations} shows high hallucination in code-review generation (~50\%) but lower in commit messages (~20\%), proposing detection metrics. Unlike these studies, we systematically measure alignment at both commit and PR levels, revealing agents are precise locally but struggle with PR-level summarization.

\section{Conclusion}
\label{sec:conclusion}
\vspace{-1.5pt}
In this paper, we compare agentic pull requests (APRs) and human pull requests (HPRs) to understand differences in code changes and PR-level communication. Analyzing 33{,}596 APRs and 6{,}618 HPRs, we find that agent-introduced symbols are removed sooner (median 3 vs.\ 34 days) and churn more (7.33\% vs.\ 4.10\%), indicating a focus on narrow tasks. Agents produce stronger commit-level messages (semantic similarity 0.72 vs.\ 0.68) but lag in PR-level summarization (PR--commit similarity 0.86 vs.\ 0.88), highlighting limited full-PR reasoning. These insights guide task assignment, review practices, and agent training for more reliable agent-assisted development.

\vspace{2pt}
\noindent\textbf{Future work.} Several opportunities remain for future work, including expanding symbol-churn analysis to additional languages using parser-based techniques, refining the labels and features used to characterize high-quality PR descriptions, and examining agent-specific behaviors to uncover differences beyond aggregate trends.

\vspace{-1.5pt}
\section*{Acknowledgment}
\vspace{-1.5pt}
This work is funded by the Natural Sciences and Engineering Research Council of Canada (NSERC): RGPIN-2025-05897.

\clearpage
\balance
\bibliographystyle{ACM-Reference-Format}
\bibliography{refs}
\end{document}